\documentclass[letterpaper]{jpsj3}
\usepackage{color}

\title{Probing a non-biaxial behavior of infinitely thin hard platelets}

\author{\name{Hitomi} \surname{Nomura}\thanks{E-mail address: hnomura@ims.ac.jp}$^{1, 2}$, \name{Tomonori \surname{Koda}$^{2}$}, \name{Akihiro \surname{Nishioka}$^{2}$} and \name{Hisashi \surname{Okumura}$^{1,3}$}
}
\inst{\address{$^{1}$Department of Theoretical and Computational Molecular Science, Institute for Molecular Science, 38 Nishigo-Naka, Myodaiji, Okazaki, Aichi 444-8585, Japan} \\
$^{2}$Graduate School of Science and Engineering, Yamagata University, 4-3-16 Jonan, Yonezawa, Yamagata 992-8510, Japan \\
$^{3}$Department of Structural Molecular Science, The Graduate University for Advanced Studies, Okazaki, Aichi 444-8585, Japan
}

\abst{We give a criterion to test a non-biaxial behavior of infinitely thin hard platelets of $D_{2h}$ symmetry based upon the components of three order parameter tensors. We investigated the nematic behavior of monodisperse infinitely thin rectangular hard platelet systems by using the criterion. Starting with a square platelet system, and we compared it with rectangular platelet systems of various aspect ratios. For each system, we performed equilibration runs by using isobaric Monte Carlo simulations. Each system did not show a biaxial nematic behavior but a uniaxial nematic one, despite of the shape anisotropy of those platelets. The relationship between effective diameters by simulations and theoretical effective diameters of the above systems was also determined.}

\kword{isotropic-nematic phase transition, isobaric Monte Carlo simulations, coexistence number density, effective diameters}

\begin{document}
\maketitle

\section{Introduction}
\begin{figure}[!th]
\begin{center}$
\begin{array}{c}
\includegraphics[scale=0.7]{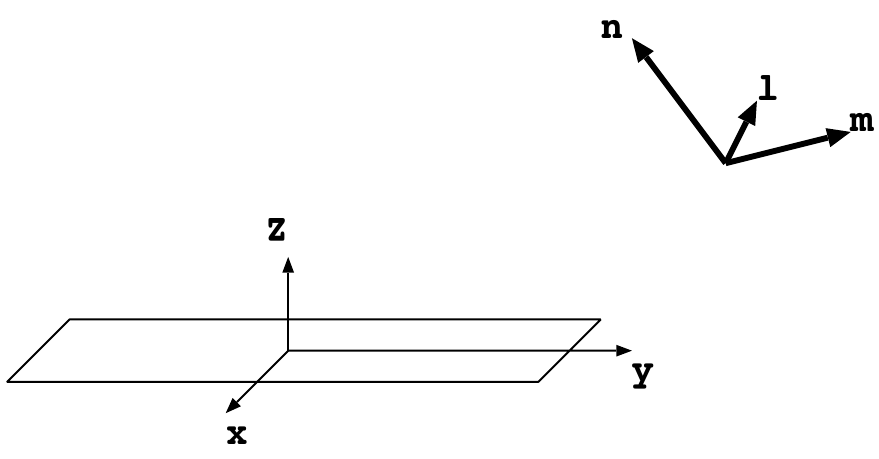}
\end{array}$
\end{center}
\caption{The first and secondary directors $\textbf{n}$ and $\textbf{m}$ by which the third $\textbf{l} \equiv \textbf{n} \times \textbf{m}$ is defined automatically. $\textbf{n}, \textbf{m}, \textbf{l}$ are called laboratory coordinate axes. On the other hand, molecular fixed coordinate axes are indicated by $\textbf{x}, \textbf{y},$ and $\textbf{z}$.}
\label{fig:fig00}
\end{figure}

\begin{figure}[!th]
\begin{center}$
\begin{array}{c}
\includegraphics[scale=0.7]{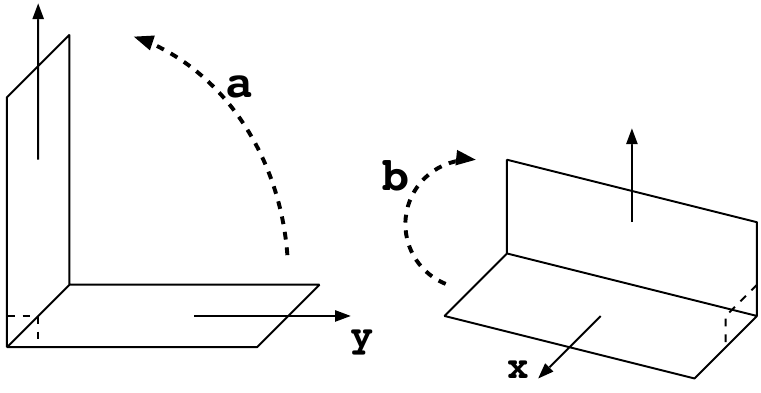}
\end{array}$
\end{center}
\caption{A model of anisotropic shaped liquid crystal molecules. The motion $\textbf{b}$ is $10$-$100$ times quicker than the motion $\textbf{a}$~\cite{tk11}.}
\label{fig:fig0}
\end{figure}

Liquid crystals are molecules with some preferred orders without any positional order. Uniaxial nematic liquid crystals have one orientational order vector, referred to as a director $\textbf{n}$. When nematic liquid crystals have not only the first director $\textbf{n}$ but also a secondary director $\textbf{m}$ which is orthogonal to $\textbf{n}$ (FIG.~\ref{fig:fig00}.), the nematic liquid crystals are said to be biaxial. They are equipped with a $10$-$100$ times quicker response time under an external field than uniaxial ones due to their anisotropic shape~\cite{k09}. (See FIG.~\ref{fig:fig0}.) In 1970, based on the molecular field theoretical calculations, Freiser predicted that anisotropic-shaped molecules should undergo a biaxial nematic phase~\cite{f07}. The first lyotropic (concentration dependent) biaxial nematic liquid crystals were synthesized by Yu and Saupe in 1980~\cite{ys08}. Twenty-four years later, the discovery of a system of thermotropic (temperature dependent) biaxial nematic liquid crystals, based on an oxadiazole unit, was announced by Maden, Dingemans et. al.~\cite{md04, dm06}, but later, it was realized that, in fact, the system consisted of a uniaxial plane of smectic C liquid crystals.~\cite{gm11} Since then, new experimental methods have been developed to detect biaxial liquid crystals, such as $^{2}$H (Deuterium) nuclear magnetic resonance~\cite{l04} and IR (Infrared)~\cite{kv10} spectroscopies, but it remains a challenge to study the biaxial properties of nematic liquid crystals in the laboratory~\cite{gm11, tk11, kv09}.

The first theoretical study of phase transitions in liquid crystals were carried out by Onsager in 1949 using hard rod and disc models~\cite{o49}. This work was extended with the advent of a Monte Carlo simulation technique~\cite{m53} by Eppenga and Frenkel in 1984 who modeled the isotropic-nematic phase transition of the system of monodisperse infinitely thin circular platelets by excluded volume effects~\cite{e84}. Bates and Frenkel detected the first order nematic-columnar phase in the system of the circular platelets~\cite{bf98} with a scaling argument introduced by Bolhuis and Frenkel.~\cite{bf6} Bates studied the isotropic-nematic phase transition of monodisperse infinitely thin circular and non-circular (regular hexagon, pentagon, square, triangle, ellipse and rectangular of a width to height aspect ratio of 1:2) platelet systems by using grand canonical Monte Carlo simulations. He pointed out that the shape of a platelet is important to determine its isotropic-nematic phase transition~\cite{b99}.  

\begin{figure}[!th]
\begin{center}$
\begin{array}{c}
\includegraphics[scale=0.7]{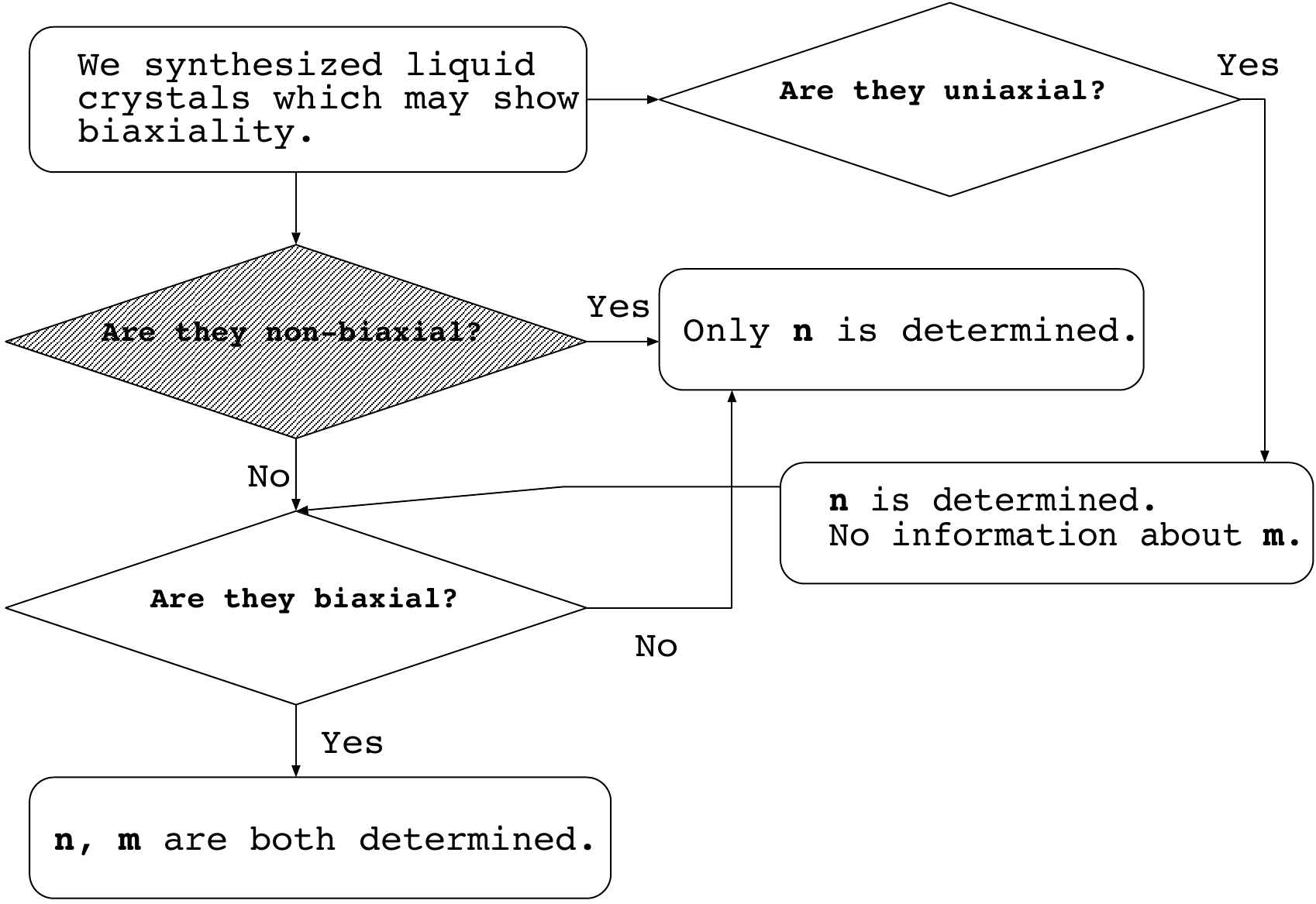}
\end{array}$
\end{center}
\caption{A flow chart for uniaxial, non-biaxial and biaxial tests. The non-biaxial test is shown by a shaded area.}
\label{fig:fig100}
\end{figure}

The plan of this paper is as follows. First, we present a set of sufficient conditions that one can use to test non-biaxiality (See FIG.~\ref{fig:fig100}) in a system of infinitely thin hard platelet-like molecules of $D_{2h}$ symmetry. Next, as a concrete example, we study the behavior of square and various rectangular platelet systems which have not been fully examined in the literature, by isobaric Monte Carlo computer simulations. Using the sufficient criterion, we check if those systems are non-biaxial. The third, we investigate how the phase transition pressure of each platelet system depends on the shape anisotropy of platelets, starting from square platelets to elongated rectangular ones.

\section{Criterion for non-biaxiality}
 For liquid crystal molecules of arbitrary shape, a generalized order parameter tensor $\mathcal{S}$ is a real symmetric (and therefore diagonalizable) matrix quantity which measures a phase transition. Its components are given by 
\begin{eqnarray}\label{eqn:S}
\mathcal{S}_{\mu \nu}^{ij} &=& \bigg\langle \mathbf{a}_{i} \cdot (\frac{3}{2} \mathbf{r}_{\mu} \otimes \mathbf{r}_{\nu} - \frac{1}{2}\mathbf{I} ) \mathbf{a}_{j} \bigg\rangle ,  \\
                          &=& \bigg\langle \frac{3}{2} (\mathbf{a}_{i} \cdot \mathbf{r}_{\mu})(\mathbf{a}_{j} \cdot \mathbf{r}_{\nu} )- \frac{1}{2}\delta_{ij}\delta_{\mu \nu} \bigg\rangle ,  \label{a}
\end{eqnarray}
where $\langle \cdot \rangle$ is an ensemble average, $\delta$ is the Kronecker delta function, $\mathbf{a}_{i, j}$ are unit vectors along three laboratory coordinate axes $\mathbf{l}, \mathbf{m}, \mathbf{n}$, and $\mathbf{r}_{\mu, \nu}$ refers to the unit vector parallel to three molecular fixed coordinate axes $\mathbf{x},\mathbf{y},\mathbf{z}$ (FIG.~\ref{fig:fig00}), and $\mathbf{I}$ is the $3 \times 3$ identity matrix~\cite{g95}. All $i, j, \mu, \nu$ take the values $1, 2$, and $3$. For the components of a nematic order parameter tensor, we set $\mu = \nu$, and then we obtain: 

 \begin{equation}\label{eqn:S1}
\mathcal{S}_{\mu\mu}^{ij}  = \bigg\langle \frac{3}{2} (\mathbf{a}_{i} \cdot \mathbf{r}_{\mu})(\mathbf{a}_{j} \cdot \mathbf{r}_{\mu} )- \frac{1}{2}\delta_{ij} \bigg\rangle.
\end{equation}
In the system of $N$ molecules, Eq.~(\ref{eqn:S1}) is computed as
 \begin{equation}\label{eqn:S2}
\mathcal{S}_{\mu\mu}^{ij}  = \bigg\langle\frac{1}{N} \sum_{k=1}^{N} \bigg\{\frac{3}{2} (\mathbf{a}_{i} \cdot \mathbf{r}_{\mu k})(\mathbf{a}_{j} \cdot \mathbf{r}_{\mu k} )- \frac{1}{2}\delta_{ij} \bigg\} \bigg\rangle.
\end{equation}
Here, the index $k$ attached to $\mathbf{r}_{\mu}$ runs over all molecules in the system.

The largest absolute value of eigenvalues of the diagonalized matrix is called a uniaxial order parameter $\mathit{S}^{\mathbf{nn}}_{\mathbf{zz}}$, and its corresponding eigenvector is the first director $\mathbf{n}$. We simplify the notation by writing as $\mathit{S}_{\mathbf{zz}} \equiv \mathit{S}^{\mathbf{nn}}_{\mathbf{zz}}, \mathit{S}_{\mathbf{yy}} \equiv \mathit{S}^{\mathbf{nn}}_{\mathbf{yy}}$ and $\mathit{S}_{\mathbf{xx}} \equiv \mathit{S}^{\mathbf{nn}}_{\mathbf{xx}}$. We can also write them in terms of Saupe parameters, $\mathsf{S}=\mathit{S}_{\mathbf{zz}}, \mathsf{D} = \mathit{S}_{\mathbf{xx}}-\mathit{S}_{\mathbf{yy}}$~\cite{dt}. To study a uniaxial behavior of anisotropic molecules, it has been often the case that one evaluates $\mathsf{D}$ by taking the difference of $\mathit{S}_{\mathbf{xx}}$ and $\mathit{S}_{\mathbf{yy}}$; however, we keep the raw values of $\mathit{S}_{\mathbf{xx}}$ and $\mathit{S}_{\mathbf{yy}}$ as they are. It is because they can tell us a non-biaxial nematic behavior of infinitely thin anisotropic molecules of $D_{2h}$ symmetry. More precisely speaking, assuming that those anisotropic molecules show a uniaxial nematic phase ($\mathit{S}_{\mathbf{zz}} > 0.7$), if we have $\mathit{S}_{\mathbf{yy}} \approx \mathit{S}_{\mathbf{xx}} \approx -0.5$, then they are in a non-biaxial phase. Supposing in stead that they were in a biaxial phase, we would have $\mathit{S}_{\mathbf{zz}} > 0.7, \mathit{S}_{\mathbf{yy}} > 0.7, \mathit{S}_{\mathbf{xx}} > 0.7$. Without any assumptions such as infinitely thin and $D_{2h}$ symmetry, we need to confirm $\mathit{S}_{\mathbf{zz}} > 0.7, \mathsf{C} > 0$ to say that arbitrary anisotropic molecules are in a biaxial nematic phase. Here, $\mathsf{C}$ is another Saupe parameter, and it is defined by $\mathsf{C}=\mathit{S}^{\mathbf{\mathbf{ll}}}_{\mathbf{xx}}+\mathit{S}^{\mathbf{\mathbf{mm}}}_{\mathbf{yy}}-\mathit{S}^{\mathbf{\mathbf{ll}}}_{\mathbf{yy}}-\mathit{S}^{\mathbf{\mathbf{mm}}}_{\mathbf{xx}}$.~\cite{dt} All numerical values used here are estimated values to describe a typical nematic phase.

The magnitude of the difference between the absolute values of the second and third largest eigenvalues $\mathit{S}_{\mathbf{zz},2},\mathit{S}_{\mathbf{zz},3}$ of the matrix $\mathcal{S}_{\mathbf{zz}}^{ij}$, which we shall call a rocking parameter, is denoted by $\mathit{R}_{\mathbf{zz}} \equiv ||\mathit{S}_{\mathbf{zz},2}| -| \mathit{S}_{\mathbf{zz},3}||$. This measures to what degree a system shows a rocking-chair like nematic behavior with two orthogonal preferred orientations. In general, the eigenvectors corresponding to the second and third largest eigenvalues are not always the second and the third director $\mathbf{m}$ and $\mathbf{l}~(=\mathbf{n} \times \mathbf{m})$. The other rocking parameters with respect to $\textbf{y},\textbf{x}$ axies $\mathit{R}_{\mathbf{yy}} \equiv ||\mathit{S}_{\mathbf{yy},2}| -| \mathit{S}_{\mathbf{yy},3}||$ and $\mathit{R}_{\mathbf{xx}} \equiv ||\mathit{S}_{\mathbf{xx},2}| -| \mathit{S}_{\mathbf{xx},3}||$ are defined similarly. Rocking parameters are relatively small; nevertheless, if a system is in a non-biaxial nematic phase, we have the following relation among them: $ \mathit{R}_{\mathbf{zz}} < \{\mathit{R}_{\mathbf{yy}}, \mathit{R}_{\mathbf{xx}}\}$. Dingemans et. al. called $\mathit{R}_{\mathbf{zz}}$ as a biaxial molecular order in contrast to a phase biaxiality $\eta = \mathit{S}^{\mathbf{\mathbf{mm}}}_{\mathbf{zz}}-\mathit{S}^{\mathbf{\mathbf{ll}}}_{\mathbf{zz}}$.~\cite{dm06} To the best of our knowledge, the meaning of $ \mathit{R}_{\mathbf{yy}}$ and $\mathit{R}_{\mathbf{xx}}$ has not been investigated yet. 

In summary, we give the above non-biaxial criteria in a comparison with uniaxial and biaxial conditions in Table~\ref{tab:a}.

\begin{table}[h]
\begin{center}
\caption{The list of phases and the set of parameters to be determined. We cannot determine that a system is non-biaxial only by checking the condition for being a uniaxial phase. The information of $\mathit{S}_{\mathbf{yy}}$ and $\mathit{S}_{\mathbf{xx}}$ are needed.}
\begin{tabular}{lll}\hline\hline
Phase & Parameter & Condition\\
\hline
Uniaxial & $\mathit{S}_{\mathbf{zz}}$ & $\mathit{S}_{\mathbf{zz}} > 0.7$\\
Non-biaxial & $\mathit{S}_{\mathbf{zz}}, \mathit{S}_{\mathbf{yy}}, \mathit{S}_{\mathbf{xx}}$ & $\mathit{S}_{\mathbf{zz}} > 0.7, \mathit{S}_{\mathbf{yy}} \approx \mathit{S}_{\mathbf{xx}} \approx -0.5$\\
Biaxial & $ \mathit{S}_{\mathbf{zz}}, \mathsf{C}$ & $\mathit{S}_{\mathbf{zz}} > 0.7, \mathsf{C} > 0$\\ \hline\hline
\\
\end{tabular}
\label{tab:a}
\end{center}
\end{table}

\section{Models}
\begin{figure}[!th]
\begin{center}$
\begin{array}{c}
\includegraphics[scale=0.8]{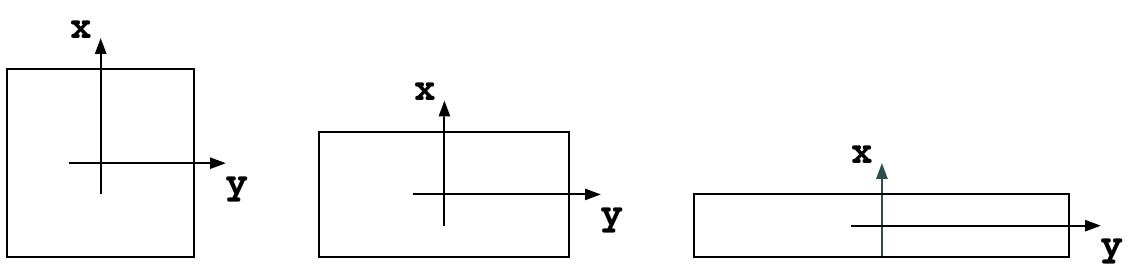}
\end{array}$
\end{center}
\caption{Square (1:1) and rectangular platelets of equal area with aspect ratios 1:2 and 1:5 are shown as examples. The $\textbf{z}$-axis is coming out of the paper where the $\textbf{x}$- and $\textbf{y}$-axes cross.}
\label{fig:fig1}
\end{figure}

Monodisperse infinitely thin rectangular (of aspect ratios 1:1, 1:2, 1:2.5, 1:3, 1:3.5, 1:4, 1:5) hard platelet systems (FIG.~\ref{fig:fig1}) are used. Previously, we examined the isotropic-nematic phase transition of an $N=120$ square (1:1) platelet system~\cite{nk12}. This time, each system is composed of $N=480$ platelets of equal area $A$. We set the areas of square and rectangular platelets to be the same. The reduced pressure is set by $p^{\ast} = D^{3}\beta p$, where the unit length $D$ is defined by $D = (\sqrt A)/2$, $\beta$ is the inverse temperature, and $p$ is a pressure. We performed equilibration runs using isobaric Monte Carlo simulations. The initial configuration of each system was isotropic. We isotropically compressed it at reduced (dimensionless) pressures $p^{\ast}$ ranged in $p^{\ast}=0.1$-$2.0$. An excluded volume repulsion drives the phase transition of each system. We compute a number density $\rho=N\sigma^3/V$, where $N$ is the number of platelets, $\sigma=\sqrt{4A/\pi}$ is a scaled diameter, and $V$ is the volume of a periodic simulation box. About up to 50000 Monte Carlo steps are typically used until equilibrium is reached. Thermodynamic quantities were calculated with up to 10000 Monte Carlo steps.

\section{Numerical results and Discussion}
\begin{figure}[]
\begin{center}$
\begin{array}{c}
\includegraphics[scale=0.7]{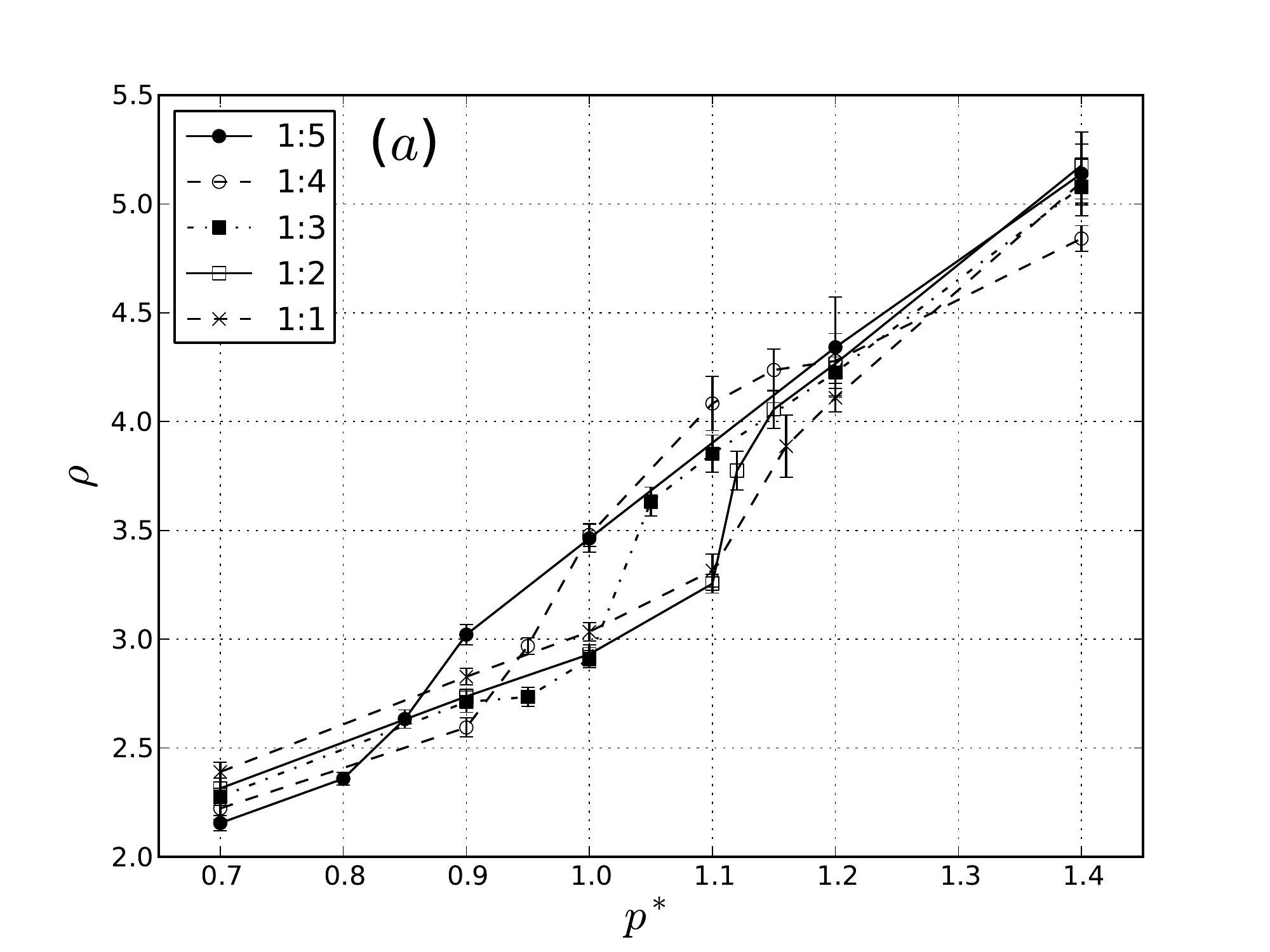}\\  
\includegraphics[scale=0.7]{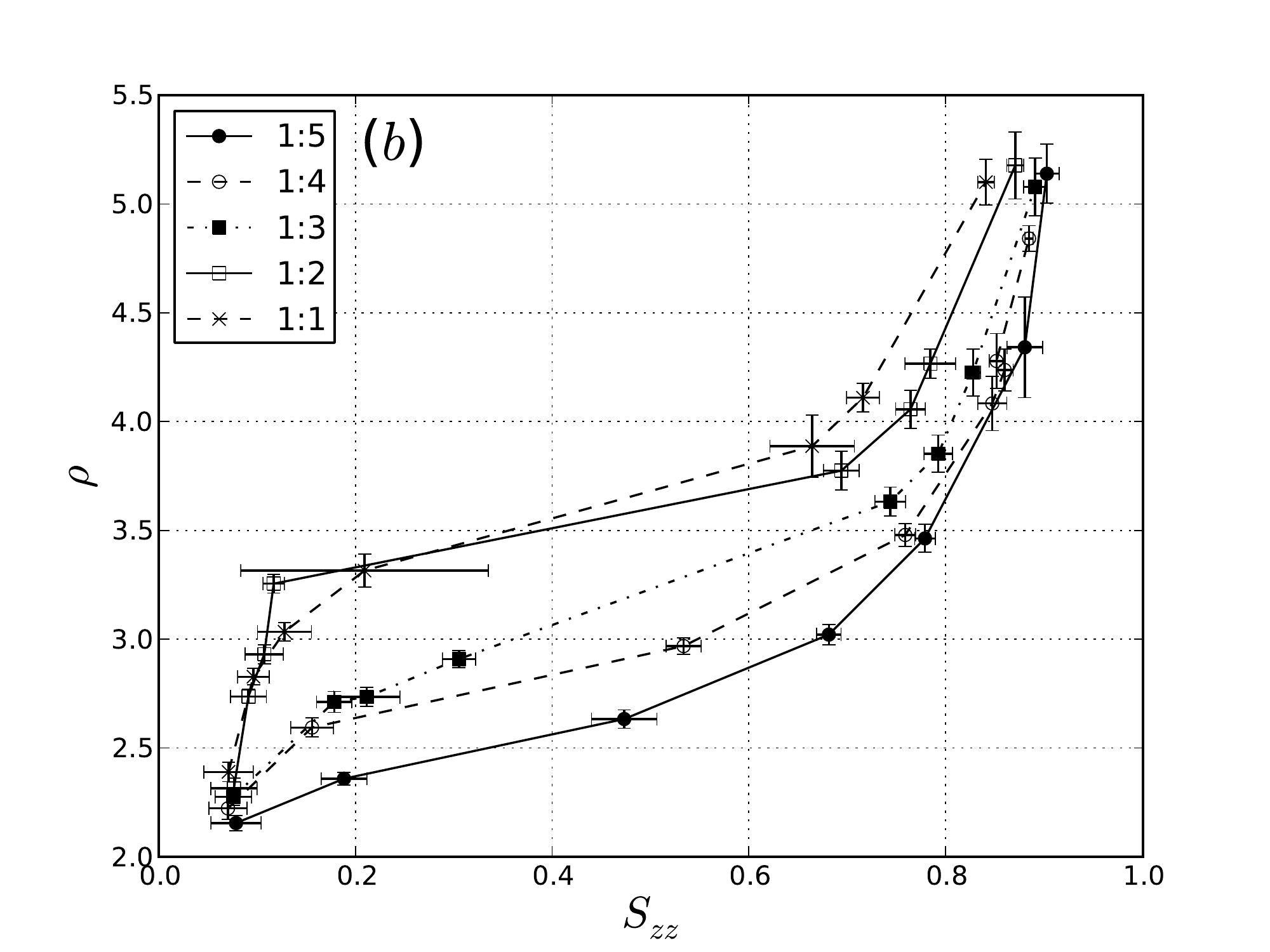} 
\end{array}$
\end{center}
\caption{(a) Pressure dependance of the number densities $\rho$. $p^{\ast}$ means a reduced pressure. (b) $\mathit{S}_{\mathbf{zz}}$ vs. $\rho$. We started from isotropic and compressed each system to nematic. Throughout this paper, unless otherwise stated, the segment lines drawn between data points are intended to guide the eye, and error bars are the standard deviation of the mean.}
\label{fig:fig4}
\end{figure}
\begin{figure}[]
\begin{center}$
\begin{array}{c}
\includegraphics[scale=0.7]{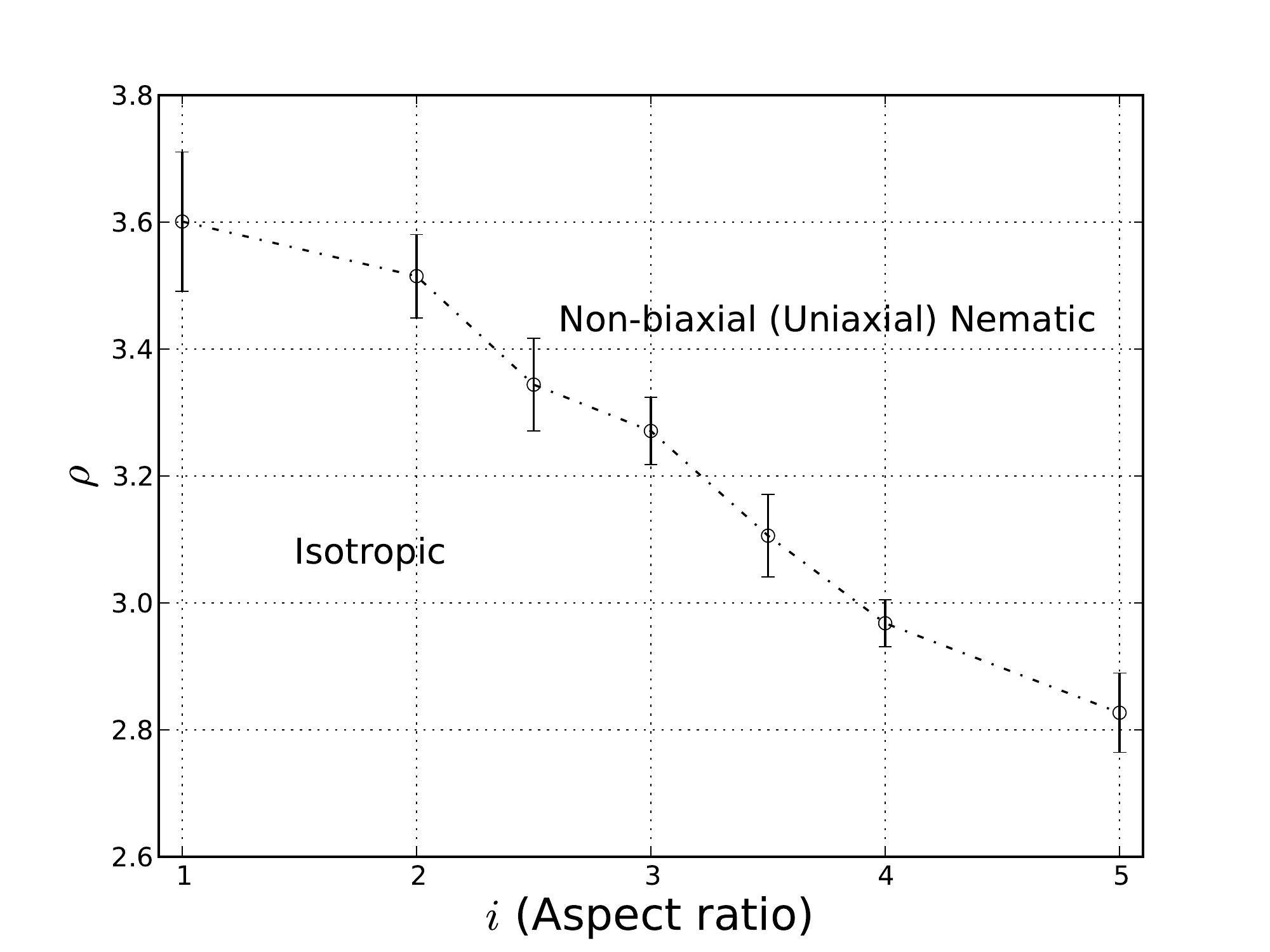}
\end{array}$
\end{center}
\caption{Phase diagram for infinitely thin rectangular hard platelets. The aspect ratio $i$ of the platelets is in the form $i=i/1$ instead of $1:i$. The dotted line drawn through the data is the phase boundary of isotropic and non-biaxial nematic.}
\label{fig:fig98}
\end{figure}

The plots of reduced pressure $p^{\ast}$ vs. the number densities $\rho$ and $\mathit{S}_{\mathbf{zz}}$ vs. $\rho$ from an isotropic phase to a nematic phase are shown in FIG.~\ref{fig:fig4}. We computed a phase transition point by taking the center of the steepest joined segment of two data points. In our previous study, the coexistence number density $\rho_{t}$ (the subscript $\textit{t}$ means a transition number density) for 120 square platelets was determined by using the phase space multi-histogram method (PSMH), and it was $\rho_{t}=3.87 \pm 0.23$ at $p^{\ast}=1.13$.~\cite{nk12} In contrast, taking the center of the steepest joined linear segment of two data points, we found that the coexistence number density of the 120 square platelet system was $\rho_{t}=3.60 \pm 0.11$ at $p^{\ast}=1.13$. The results from both methods corresponded well within errors. 
As an aspect ratio increased, a phase transition pressure (or coexistence number density) decreased. In Table~\ref{tab:b} and FIG.~\ref{fig:fig98}, we summarized phase transition reduced pressures and coexistence densities and the phase diagram for platelets of various aspect ratios. 

\begin{table}[h]
\begin{center}
\caption{Phase transition reduced pressures ($p^{\ast}$) and coexistence number densities ($\rho_{t}$) of platelets of various aspect ratios.}
\begin{tabular}{lll}\hline\hline
Aspect ratio & $p^{\ast}$ & $\rho_{t}$\\
\hline
1:1 & 1.13 & 3.60 $\pm$ 0.11\\
1:2 & 1.11 & 3.52 $\pm$ 0.07\\
1:2.5 & 1.05 & 3.34 $\pm$ 0.07\\
1:3 & 1.03 & 3.27 $\pm$ 0.05\\
1:3.5 & 0.98 & 3.11 $\pm$ 0.06\\
1:4 & 0.95 & 2.97 $\pm$ 0.04\\
1:5 & 0.88 & 2.83 $\pm$ 0.06\\ \hline\hline
\\
\end{tabular}
\label{tab:b}
\end{center}
\end{table}

\begin{figure}[]
\begin{center}$
\begin{array}{c}
\includegraphics[scale=0.7]{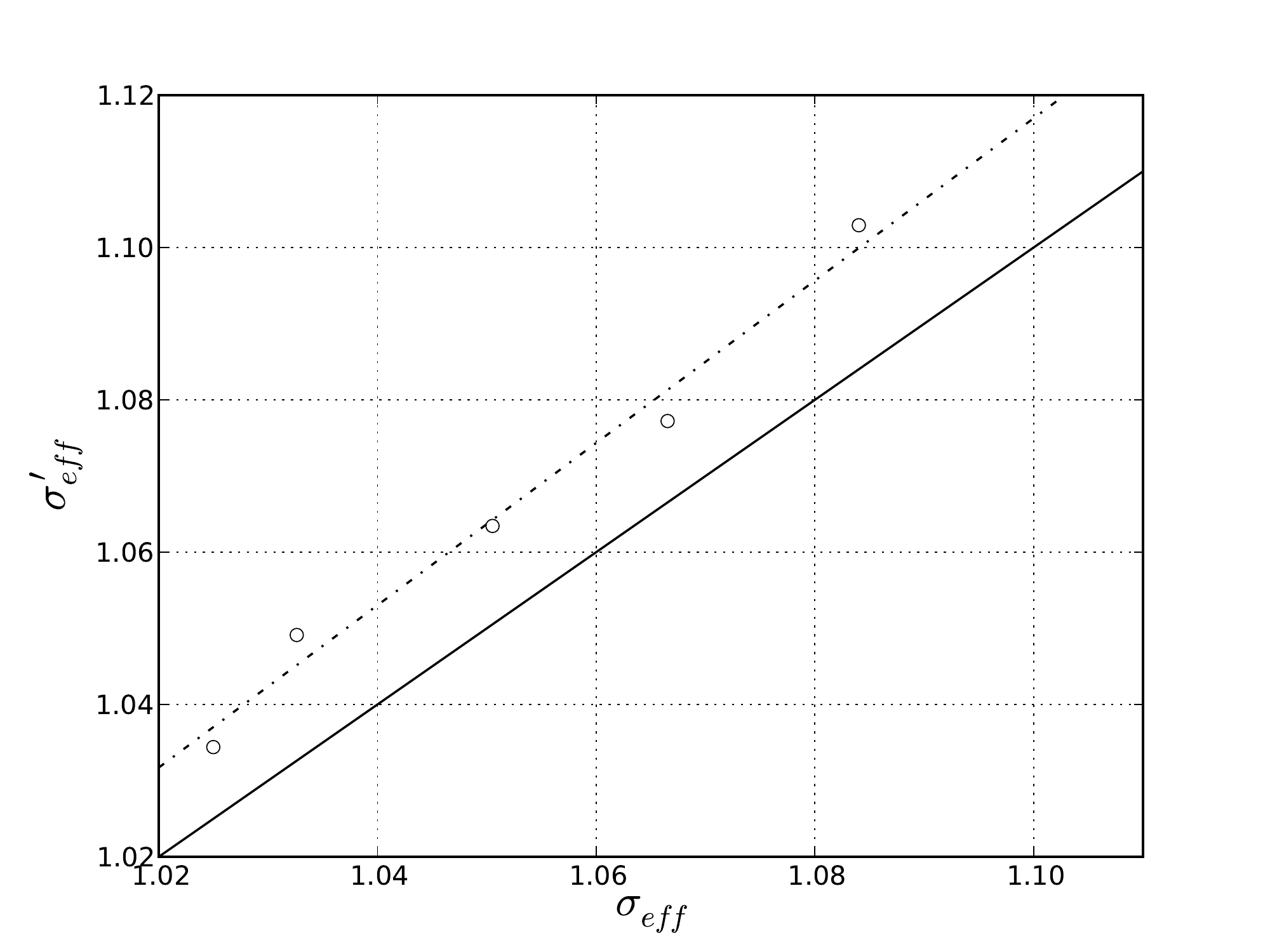}
\end{array}$
\end{center}
\caption{Effective diameters from simulations vs. theoretically calculated effective diameters. The solid line indicates $\sigma_{\text{eff}} = \sigma_{\text{eff}}^{'}$, and the obtained data was fitted by the dotted line $1.07\sigma_{\text{eff}}- 0.06 = \sigma_{\text{eff}}^{'}$.}
\label{fig:fig983}
\end{figure}

Before going into the details of our further results, we shall recall the definition of effective diameters. Let us denote the number densities of the system of rectangular platelets by $\rho_{\alpha i}$. The lower index $i$ adjacent to $\alpha$ indicates the aspect ratio of a rectangular platelet $\alpha$ written in the form $i=i/1$ instead of $1:i$. As Bates defined an effective diameter to be the cube root of the ratio of the isotropic-nematic coexistence number density for a non-circular platelet system to a circular platelet system~\cite{b99}, we analogously define by $\sigma_{\text{eff}}=\sqrt[3]{\rho_{\alpha 1}/\rho_{\alpha i}}$. Here, $\rho_{\alpha 1}, \rho_{\alpha i}$ are coexistence densities, and strictly written, it might be denoted as $\rho_{t, \alpha i}$. For simplicity, we adapt the notation with omitting the subscript $t$ indicating a phase transition as before.  Following Kihara's work~\cite{k53}, Bates computed the orientation ensemble average excluded volume (denoted by $\langle V_{\alpha i}^{\text{ex}}\rangle$ in this paper) of a pair of infinitely thin platelets of arbitrary convex shape and found that $\langle V_{\alpha i}^{\text{ex}}\rangle = \frac{1}{2}AP_{\alpha i}$, where $A$ is the area of a platelet $\alpha i$, and $P_{\alpha i}$ is its perimeter.~\cite{b99}  In the same spirit as Bates', taking the cube root of the ratio of excluded volumes, we define a theoretical effective diameter $\sigma_{\text{eff}}^{'}=\sqrt[3]{\langle V_{\alpha i}^{\text{ex}}\rangle/\langle V_{\alpha 1}^{\text{ex}}\rangle}$. It is of use to check whether $\sigma_{\text{eff}} = \sigma_{\text{eff}}^{'}$ holds or not since if so, one can predict systems of other aspect ratios by computing the orientation average excluded volume of a pair of platelets without performing further simulations. In an earlier study, it was expected that  $\sigma_{\text{eff}} = \sigma_{\text{eff}}^{'}$ will not hold for rectangular platelets as their aspect ratio is bigger than 1:2 since circular platelets were taken instead of square platelets (See FIG.~\ref{fig:fig1}.)~\cite{b99} We see clearly in FIG.~\ref{fig:fig983} that $\sigma_{\text{eff}} = \sigma_{\text{eff}}^{'}$ did not hold, but there observed that the measured effective diameters by simulations were smaller than the theoretical ones. The fitted line had its slope $\sigma_{\text{eff}}^{'}/\sigma_{\text{eff}} \approx 1.07$.

\begin{figure}[h]
\begin{center}$
\begin{array}{ccc}
\includegraphics[scale=0.5]{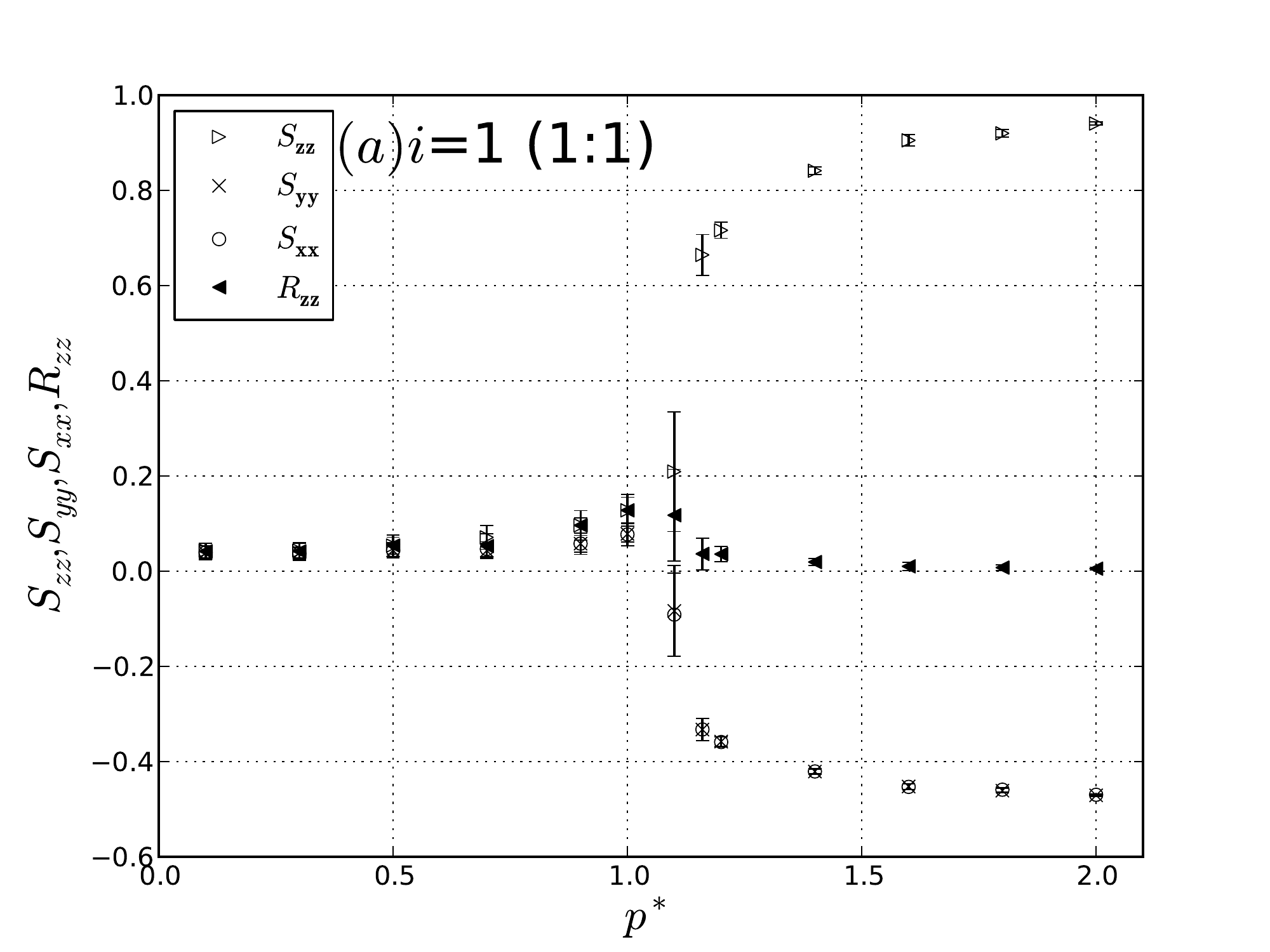}  &\\
\includegraphics[scale=0.5]{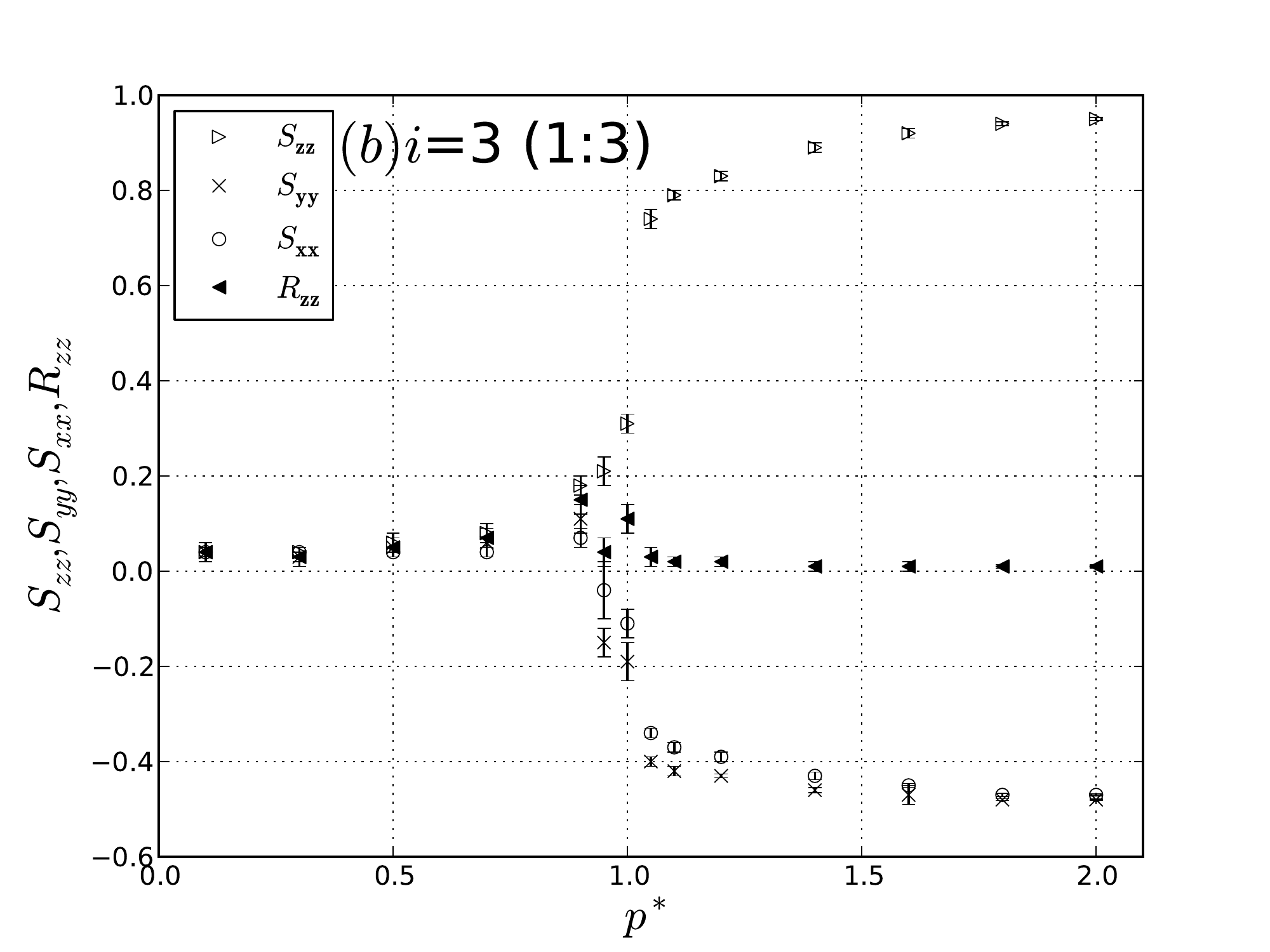}  &\\
\includegraphics[scale=0.5]{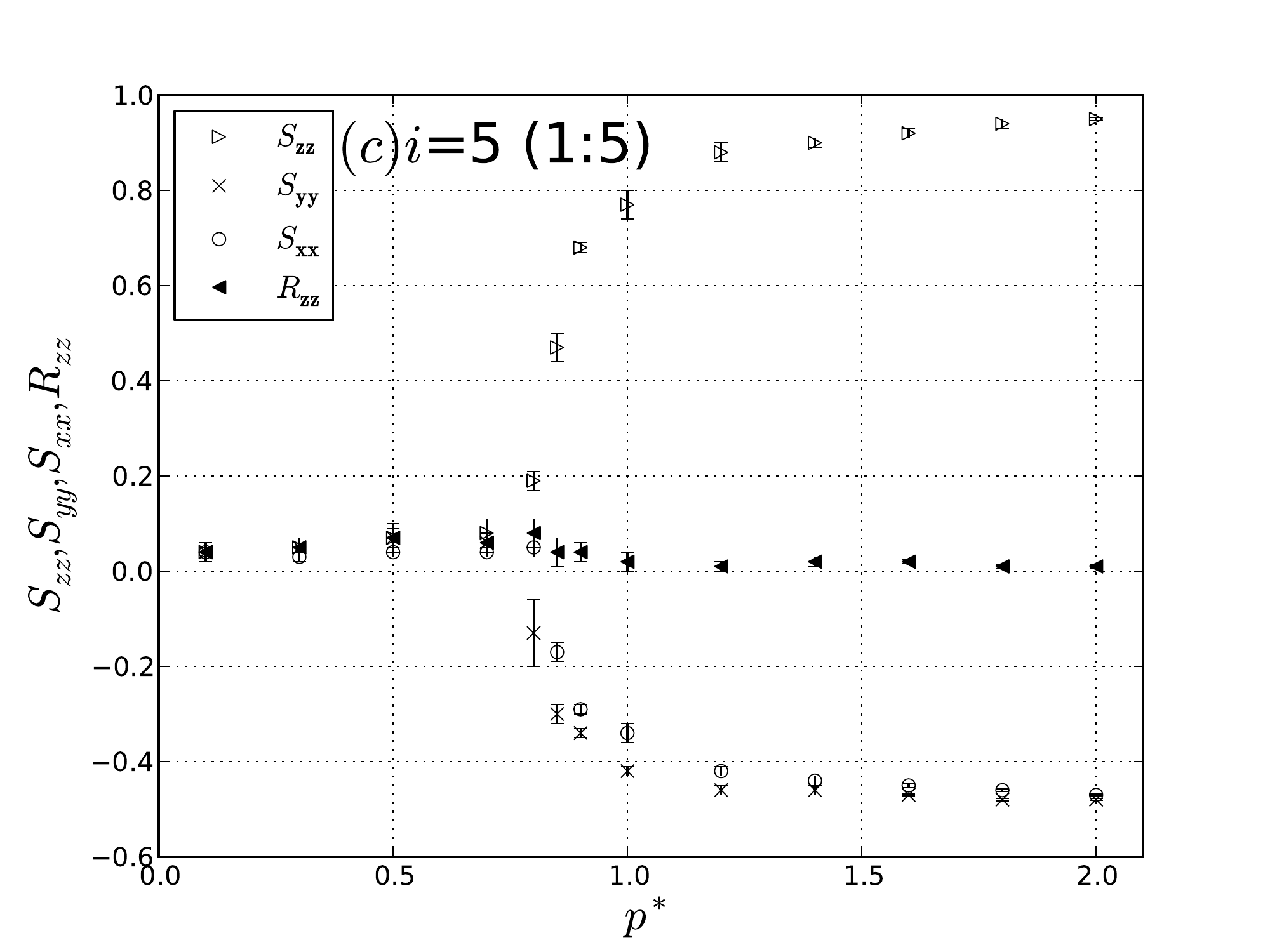} 
\end{array}$
\end{center}
\caption{Pressure dependance of the order parameters $\mathit{S}_{\textbf{zz}}, \mathit{S}_{\textbf{yy}}, \mathit{S}_{\textbf{xx}}$ and the rocking parameter $\mathit{R}_{\textbf{zz}}$ in the systems of 480 rectangular platelets of aspect ratios (top) 1:1, (middle) 1:3 and (bottom) 1:5.}
\label{fig:fig5}
\end{figure}

\begin{figure}[h]
\begin{center}$
\begin{array}{ccc}
\includegraphics[scale=0.47]{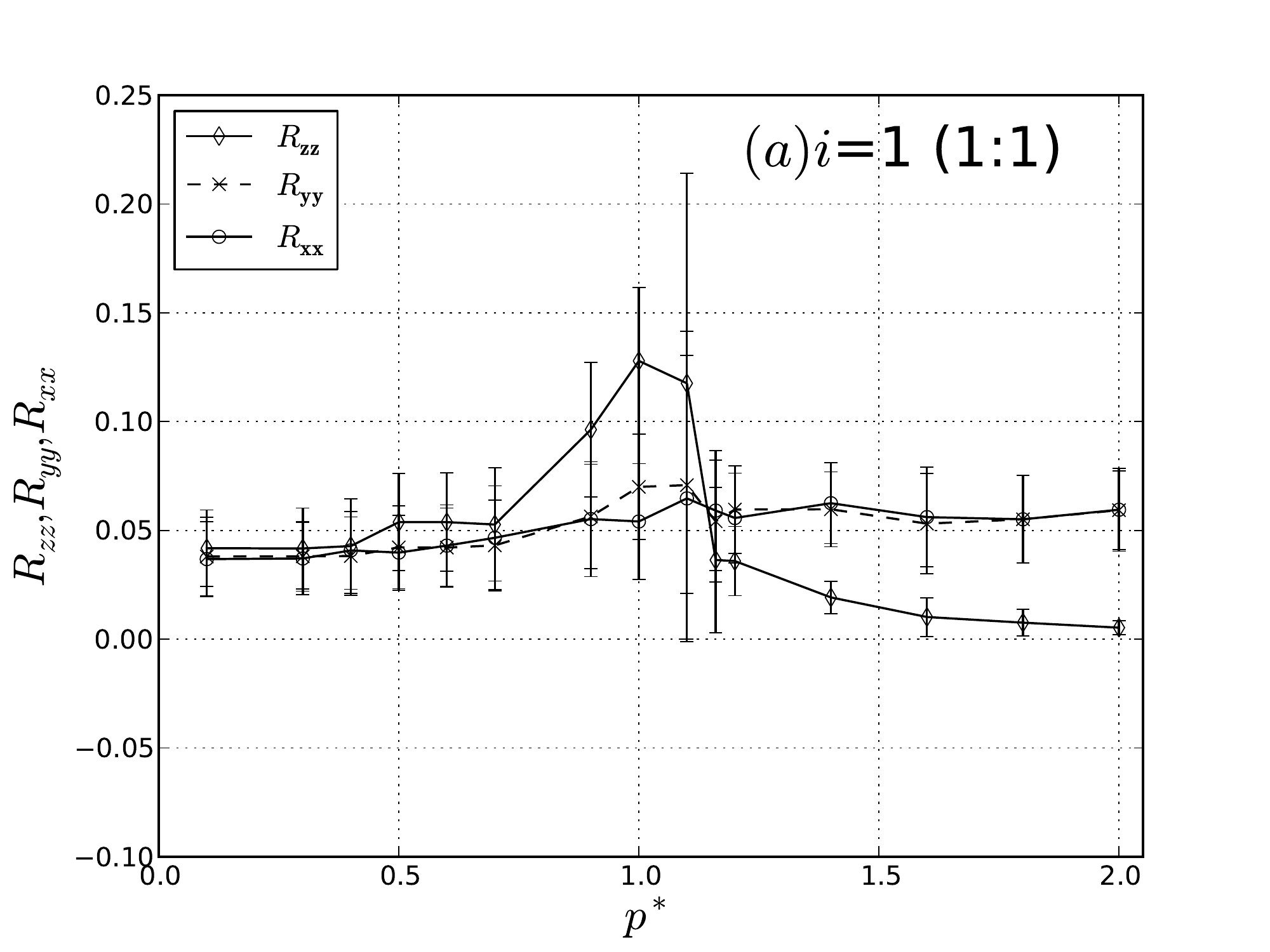}  &\\
\includegraphics[scale=0.47]{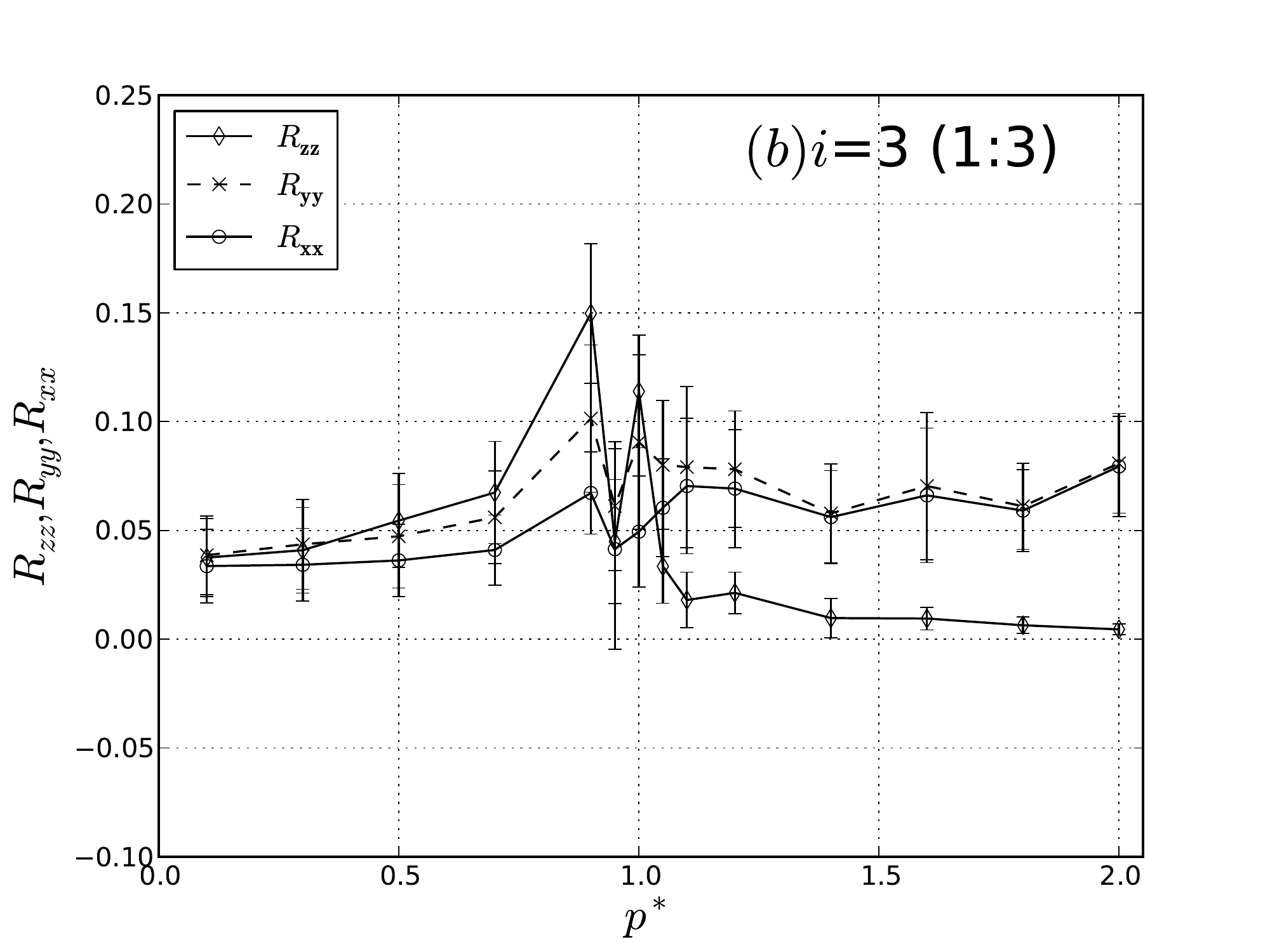}  &\\
\includegraphics[scale=0.47]{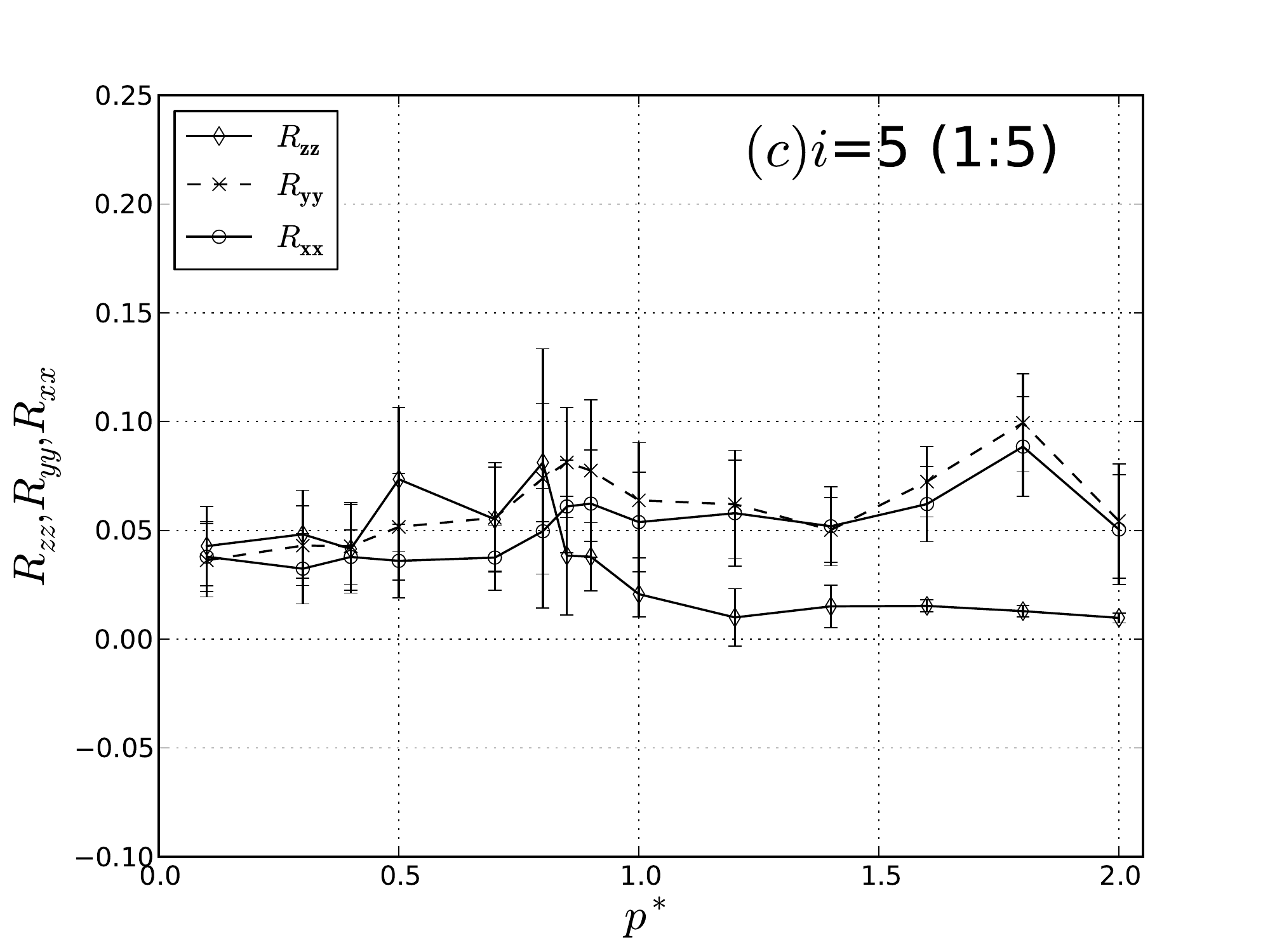}  

\end{array}$
\end{center}
\caption{$p^{\ast}$ vs. $\mathit{R}_{\textbf{zz}}, \mathit{R}_{\textbf{yy}}, \mathit{R}_{\textbf{xx}}$. (a) 1:1, (b) 1:3 and (c) 1:5.}
\label{fig:fig7}
\end{figure}

\begin{figure}[h]
\begin{center}$
\begin{array}{ccc}
\includegraphics[scale=0.4]{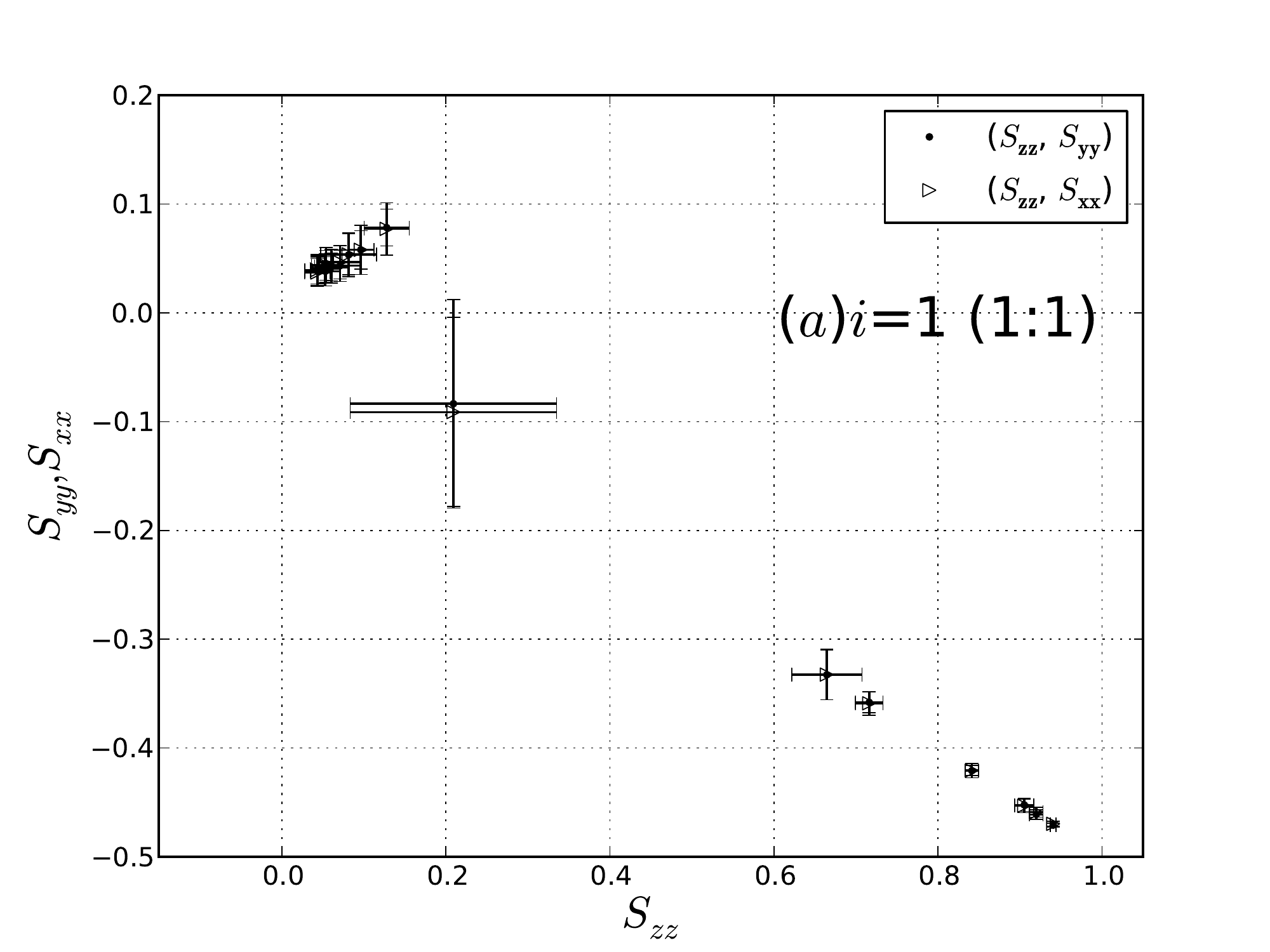}  &\\
\includegraphics[scale=0.4]{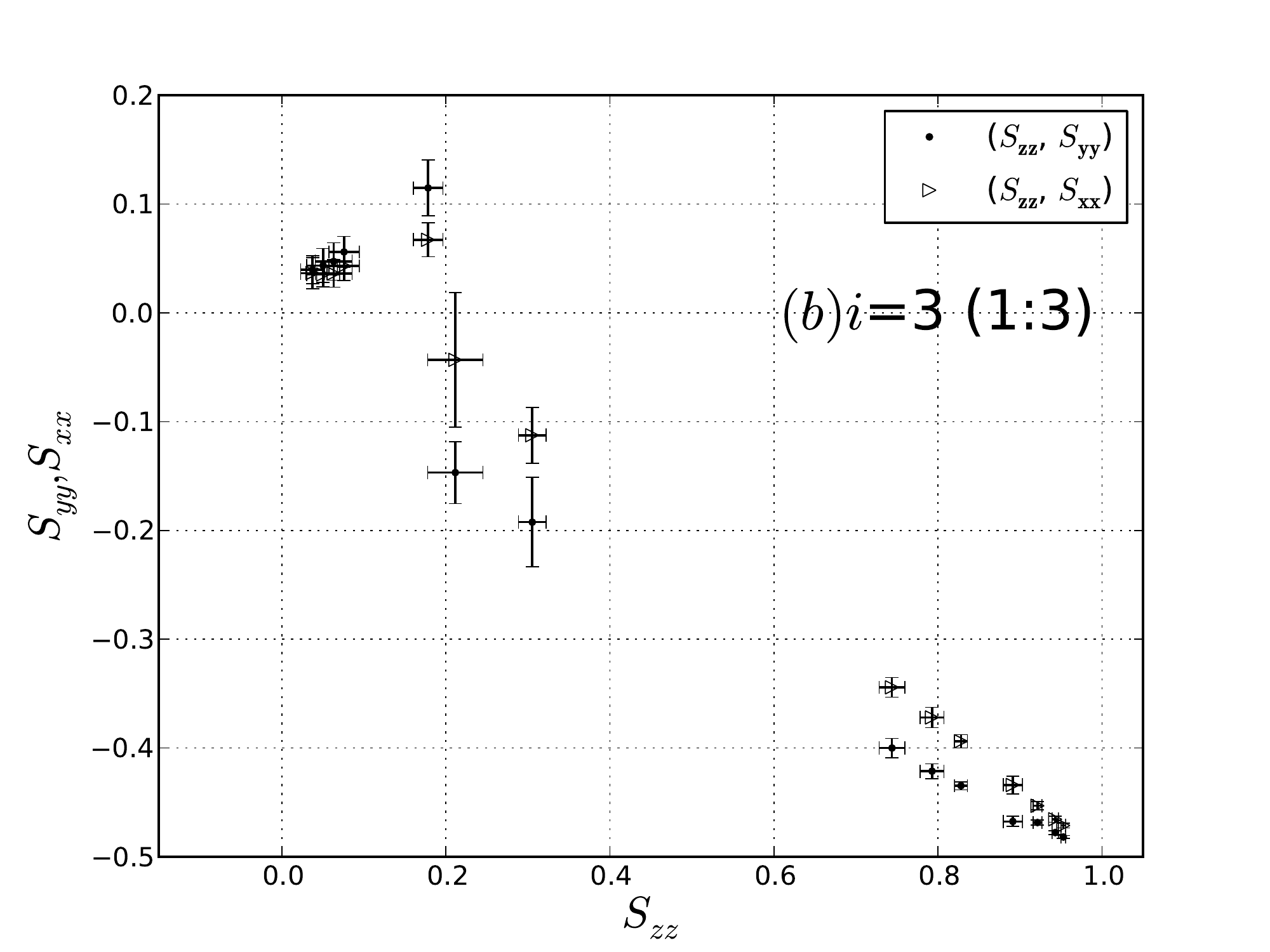}  &\\
\includegraphics[scale=0.4]{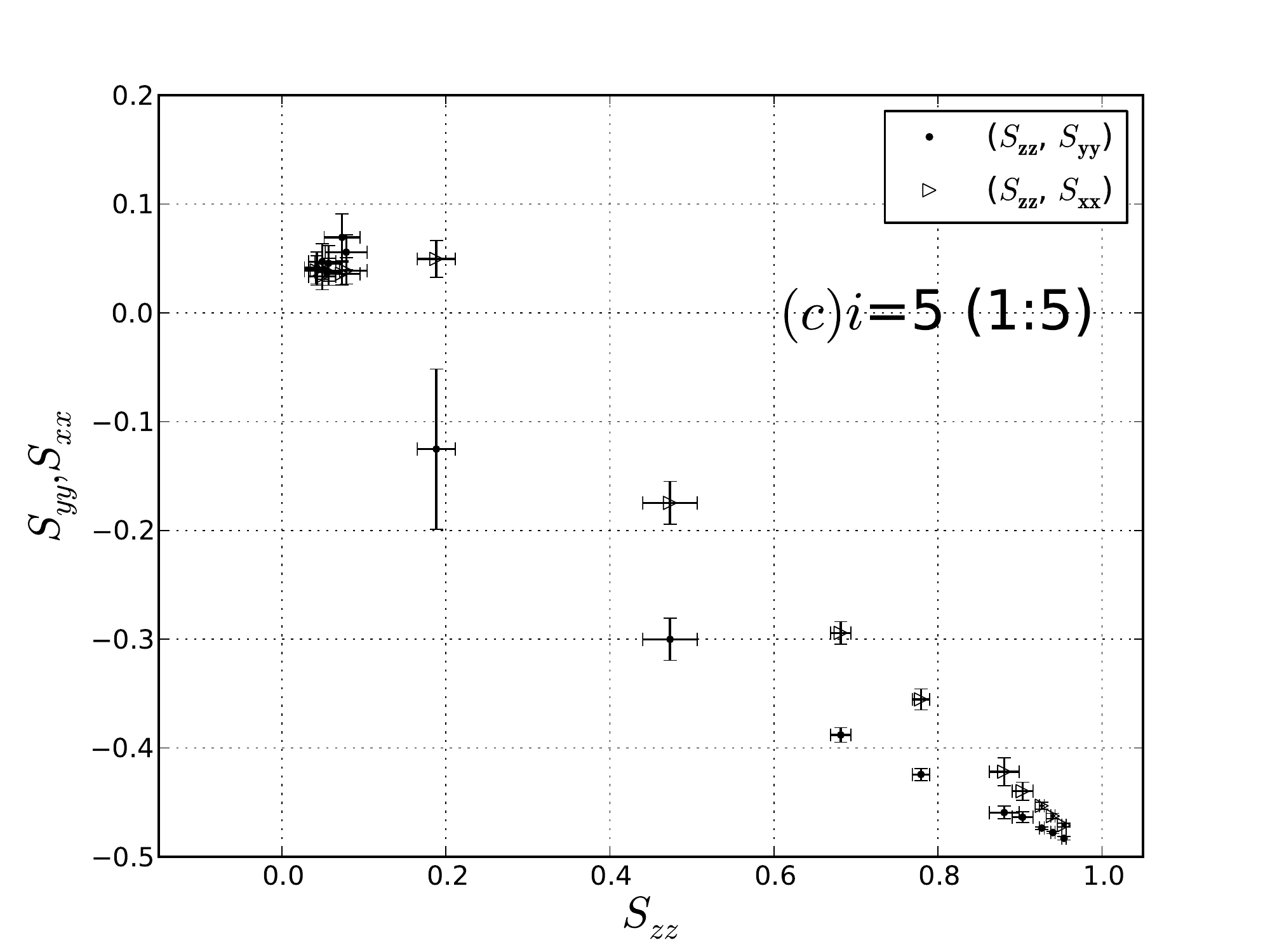}
\end{array}$
\end{center}
\caption{ $\mathit{S}_{\textbf{zz}}$ vs.  $\mathit{S}_{\textbf{yy}}$ or $\mathit{S}_{\textbf{xx}}$. (a) 1:1, (b) 1:3 and (c) 1:5.}
\label{fig:fig8}
\end{figure}

\begin{figure}[h]
\begin{center}$
\begin{array}{c}
\includegraphics[scale=0.7]{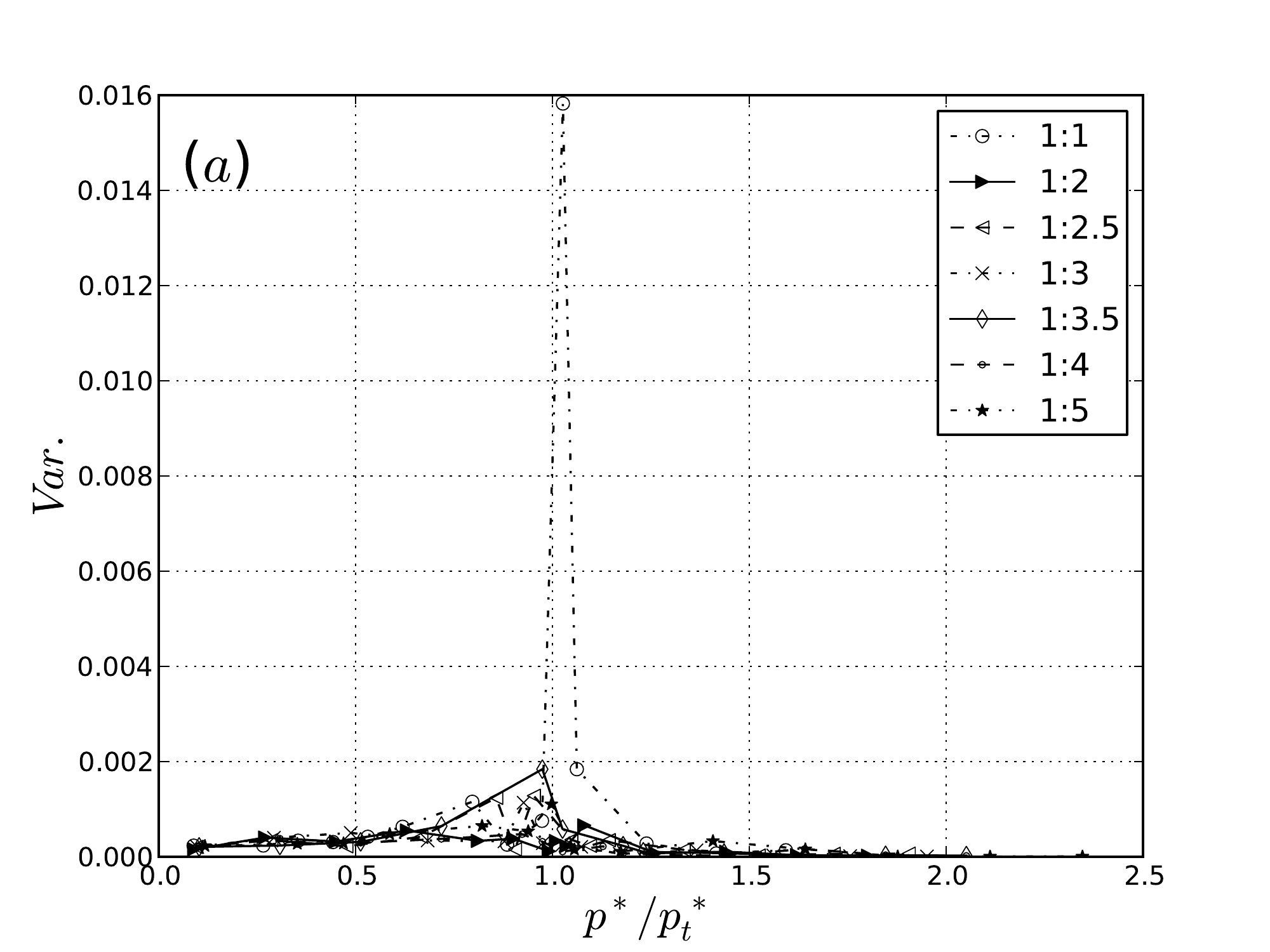}\\
\includegraphics[scale=0.7]{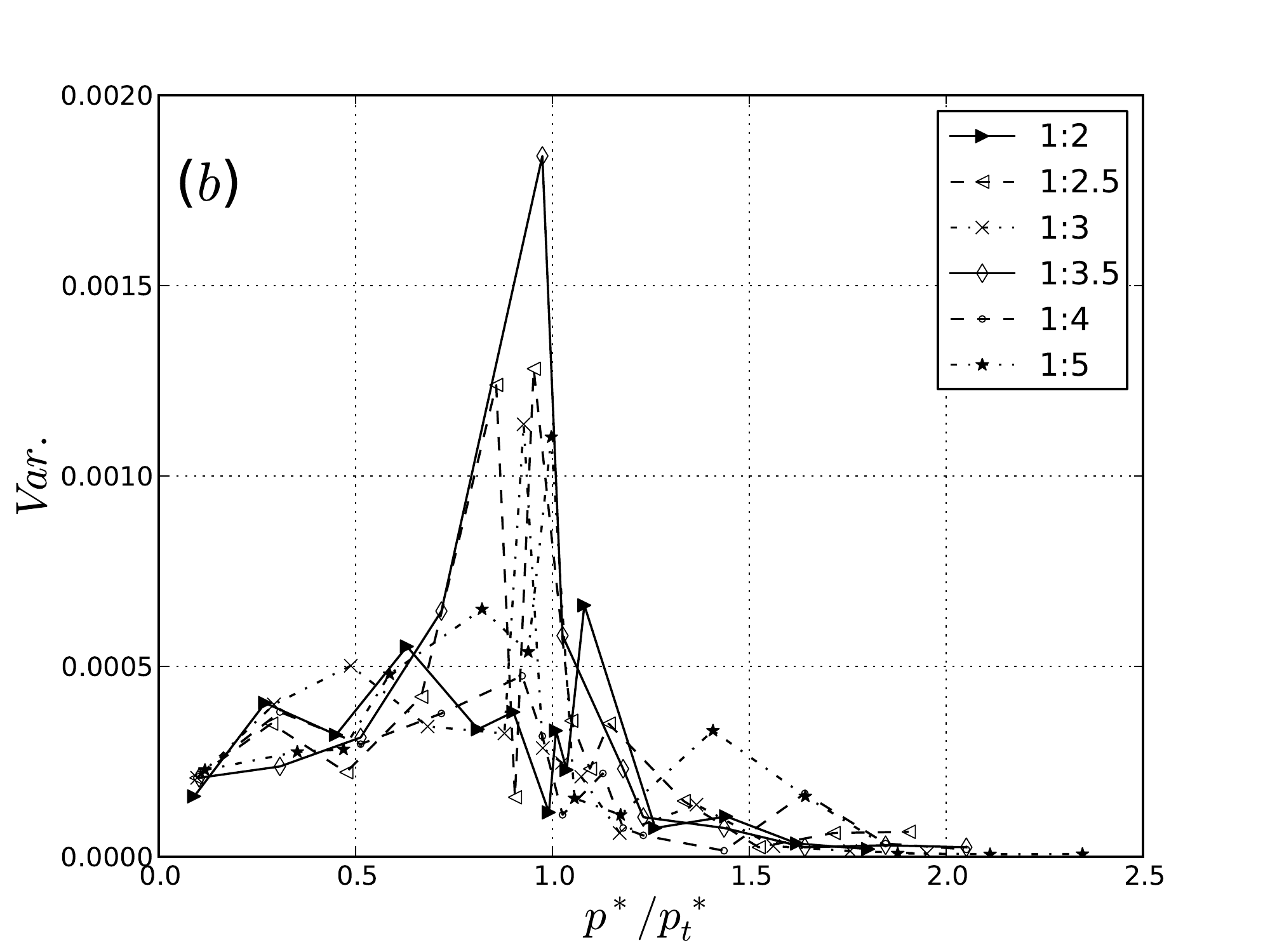}
\end{array}$
\end{center}
\caption{(a) Variances (abbreviated by Var.) of $\mathit{S}_{\textbf{zz}}$, written more precisely $\langle \delta \mathit{S}_{\textbf{zz}}^2 \rangle$, with $p^{\ast}/p^{\ast}_{t}$ for various aspect ratios. Here, $p^{\ast}_{t}$ is the reduced pressure of the isotropic-nematic phase transition. (b) The magnified plot of (a).}
\label{fig:fig99}
\end{figure}

\begin{figure}[h]
\begin{center}$
\begin{array}{c}
\includegraphics[scale=0.7]{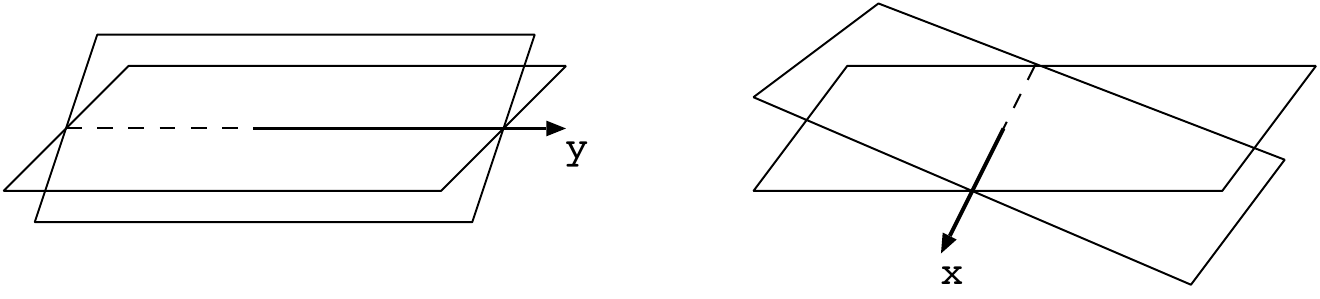}
\end{array}$
\end{center}
\caption{Swinging platelets along the $\textbf{y}$ and $\textbf{x}$-axes.}
\label{fig:fig9}
\end{figure}

FIG.~\ref{fig:fig5} displays $p^{\ast}$ vs. the order parameters $\mathit{S}_{\textbf{zz}}, \mathit{S}_{\textbf{yy}}, \mathit{S}_{\textbf{xx}}$ and the rocking parameter with respect to $\textbf{z}$-axis $\mathit{R}_{\textbf{zz}}$ of the systems of rectangular platelets of aspect ratios 1:1, 1:3 and 1:5. (Those for other aspect ratios are not shown here.) Following the $\mathit{S}_{\textbf{zz}}$ plot, the first order phase transition pressure gradually became low as the aspect ratio increased. The non-biaxial (uniaxial) nematic phase for square and rectangular platelet systems was detected from $\mathit{S}_{\textbf{zz}} > 0.7, \mathit{S}_{\mathbf{yy}} \approx \mathit{S}_{\mathbf{xx}} \approx -0.5$.  We found a low rocking parameter $\mathit{R}_{\textbf{zz}} < 0.25$ for every shape. A physical meaning of the values of $\mathit{R}_{\textbf{yy}}$ (or $\mathit{R}_{\textbf{xx}}$) is that platelets swing like a rocking chair as they rotate around each $\textbf{z}$-axis.
In FIG.~\ref{fig:fig7}, we show $p^{\ast}$ vs. $\mathit{R}_{\textbf{zz}}, \mathit{R}_{\textbf{yy}}, \mathit{R}_{\textbf{xx}}$ of the systems, and we found the relationship $ \mathit{R}_{\mathbf{zz}} < \{\mathit{R}_{\mathbf{yy}}, \mathit{R}_{\mathbf{xx}}\}$. We also found that $\mathit{S}_{\textbf{zz}} > 0.7$ and $\mathit{S}_{\textbf{yy}} < 0$ (or $\mathit{S}_{\textbf{xx}} < 0$) from FIG.~\ref{fig:fig8}. Among the systems, fluctuations (variations in  $\mathit{S}_{\textbf{zz}}$) of lower aspect ratios were bigger than higher ones near their phase transition points. (See FIG.~\ref{fig:fig99}) These indicated that the more elongated platelets showed a uniaxial nematic phase as they were rotating around the $\textbf{z}$-axis in the more stable manner, and in addition, they swung along the $\textbf{y}$- and $\textbf{x}$-axes. (See FIG.~\ref{fig:fig9}.) This could be observed simply by computing the components of order parameter tensors with respect to the $\textbf{z}$, $\textbf{y}$, and $\textbf{x}$-axes. If it were in a biaxial nematic phase, we would expect to have $\mathit{S}_{\textbf{zz}} > 0.7, \mathit{S}_{\textbf{yy}} > 0.7, \mathit{S}_{\textbf{xx}} > 0.7$, and $\mathit{R}_{\textbf{zz}} \approx \mathit{R}_{\textbf{xx}} \approx \mathit{R}_{\textbf{xx}} \approx 0$.

\section{Conclusions}
We introduced a non-biaxiality criterion for infinitely thin anisotropic molecules of $D_{2h}$ symmetry. This is a test for uniaxiality, but it cannot be applied to prove biaxiality. To confirm biaxiality, we would need to add additional constraints on the $\mathsf{C}$ parameter which can be determined~\cite{dt} with additional simulations. Our criteria for non-biaxiality offers an efficient means to rule out biaxiality. We studied the isotropic-nematic behavior of monodisperse infinitely thin square and rectangular hard platelet model systems by isobaric Monte Carlo simulations. Although one might expect that the shape anisotropy would generate a biaxial nematic phase, we demonstrated the shape anisotropy of the platelet systems, from square platelets to rectangular ones, did not induce a biaxial nematic behavior by calculating components of order parameter tensors. Those platelets were in a non-biaxial (uniaxial) nematic phase at a high reduced pressure. We observed that they were rotating around the  $\textbf{z}$-axis as they swung along the $\textbf{y}$- and $\textbf{x}$-axes. The sufficient criterion to check systems of monodisperse infinitely thin liquid crystal molecules of $D_{2h}$ symmetry being non-biaxial was presented. The relationship seen in effective diameters of those systems was determined.

\end{document}